\documentclass[prl,aps,showpacs,nofootinbib,twocolumn]{revtex4}
\usepackage{epsfig}
\usepackage{amssymb}
\usepackage{amsmath}
\usepackage{amsfonts}
\begin{document}
\title{A new effective interaction for the trapped Fermi gas}
\author{Y. Alhassid,$^{1}$  G.F. Bertsch,$^{2}$ and L. Fang$^{1}$}
\affiliation{$^{1}$Center for Theoretical Physics, Sloane Physics
Laboratory, Yale University, New Haven, CT 06520\\
$^{2}$Department of Physics and Institute of Nuclear Theory,
Box 351560\\ University of Washington, Seattle, WA 98915}
\date{June 27, 2007}
\def\lb{\langle}
\def\rb{\rangle}
\def\ni{\noindent}
\def\be{\begin{equation}}
\def\ee{\end{equation}}
\def\sumk{\sum_k}
\def\ad{a^\dagger_k}
\def\adb{a^\dagger_{\bar k}}
\def\a{a_k}
\def\ab{a_{\bar k}}
\def\Tr{{\rm Tr}}

\begin{abstract}
We apply the configuration-interaction method to calculate the spectra of
two-component Fermi systems in a harmonic trap, studying the convergence of
the method at the unitary interaction limit. We find that for a fixed
regularization of the two-body interaction the convergence is exponential or
better in the truncation parameter of the many-body space.  However, the
conventional regularization is found to have poor convergence in the
regularization parameter, with an error that scales as a low negative power of this
parameter.  We propose a new regularization of the two-body interaction that
produces exponential convergence for systems of three and four particles.
From the systematics, we estimate the ground-state energy of the four-particle system
to be $(5.05\pm0.024)\hbar \omega$.
\end{abstract}


\maketitle

The study of cold trapped atomic condensates has become a rich field
experimentally. By providing a strongly interacting system that is well
defined, it also offers physicists an unprecedented opportunity to assess
theoretical techniques that cross the boundaries of disciplines.  In the
so-called unitary limit, the only dimensional scale of the problem is fixed
by the harmonic trap frequency.  Systematic studies have begun on small
systems using fixed-node Monte Carlo~\cite{ch07,st07} and density functional
methods~\cite{bu07}. Remarkably, the exact wave functions and energies
of the $A=3$ system are known, calculated by solving a single transcendental
equation~\cite{we06}.
Our work here is in the context of the
configuration-interaction (CI) method, widely used in atomic, molecular, and
nuclear spectroscopy. We study the convergence of the CI method with respect
to a regularization parameter of the two-body interaction and find that a
simple regularization scheme that renormalizes the interaction produces slow
convergence of the three- and four-particle spectra. We introduce a new
effective interaction that gives exponential convergence, at least in
small systems.

\noindent {\em Hamiltonian.}
The cold trapped atom system is modeled by the Hamiltonian
\be\label{hamiltonian}
H = -\sum_{i=1}^A {\hbar^2 \over 2 m} \nabla_i^2 + \sum_{i=1}^A \frac{1}{2} m \omega^2 r_i^2 +\sum_{i <j} V_0 \delta(\bf r_i -\bf r_j) \;,
\ee
where $A$ is the number of atoms, $\omega$ is the trap frequency, and $V_0$
is the interaction strength.  We have two-component fermionic systems in
mind, which controls the symmetry of the allowed states.  The interaction
is represented as a $\delta$ function (contact interaction) but as we
shall see below it requires a regularization. Here we focus on an attractive
contact interaction in the unitary limit of infinite scattering length.

\noindent {\it The two-particle problem.}
The two particle system ($A=2$) is separable in center of mass and relative
coordinates $\bf r =\bf r_2 -\bf r_1$. The center of mass Hamiltonian
 describes an harmonic oscillator with frequency $\omega$ and
mass $2m$, while the relative-coordinate Hamiltonian is
$H_{\rm rel} = -{\hbar^2 \over 2 \mu} \nabla_{\bf r}^2
  + \frac{1}{2}\mu \omega^2 r^2 +V_0 \delta({\bf r})$
with reduced mass $\mu=m/2$.  The two-particle energies are given by
\be\label{2p}
E= (2{\cal N} + {\cal L} +3/2) \hbar \omega + \varepsilon_{nl},
\ee
where ${\cal N}, {\cal L}$ and $n,l$ are the radial quantum number and angular
momentum of the center of mass and relative motion, respectively.
The energies $\varepsilon_{nl}$ are the eigenvalues of $H_{\rm rel}$, and may be derived from
the boundary condition at the origin imposed by the unitary
interaction~\cite{p2spectrum}.
The contact interaction affects only the $l=0$
partial waves, and shifts each $s$-wave oscillator energy down by one
 unit of $\hbar\omega$~\cite{bu98}. Thus we have
\be\label{relative-eigen}
\varepsilon_{nl} = (2n + l +3/2 -\delta_{l,0}) \hbar \omega
\;;\;\;\;\;n=0,1,2,\ldots\;
\ee

\noindent {\em The renormalized contact interaction.}
In the CI method, the contact interaction
in Eq.~(\ref{hamiltonian}) must be treated explicitly.  However, a $\delta$-function
interaction cannot be used in three dimensions without a regularization.
We shall do this by truncating the space of relative-coordinate wave functions
to a $q$ subspace defined by the lowest $q+1$ oscillator $l=0$
wave functions (see also Ref.~\cite{ste07}).
Within the truncated space the relative-coordinate Hamiltonian can
be written as
\be\label{ho_basis}
\!\!\!\!(H_{\rm rel})^{(q)}_{n,n'} = (2 n + 3/2) \hbar\omega \delta_{n,n'} + V^{(q)}_{n,n'}
\;\;(0\leq n,n'\leq q)
\ee
where
\be\label{truncated-delta}
V^{(q)}_{n,n'} = \hbar \omega \chi_q\psi_n(0)\psi_{n'}(0) \;,
\ee
and $\psi_n(0)= \pi^{-3/4}\sqrt{(2n+1)!!/( 2^n n!)}$ is the $(n,l=0)$
oscillator wave function at $r=0$ for an oscillator of radius 1. The parameter $\chi_q$ is a dimensionless normalization constant related to $V_0$ by $\chi_q = (\hbar^2/\mu)^{-3/2}
(\hbar\omega)^{1/2}V_0$.

We determine the normalization constant $\chi_q$ by requiring the
ground-state energy of the truncated Hamiltonian to equal the exact value
for the unitary contact interaction, $\varepsilon_{00}=
\hbar\omega/2$.   The separable form of
(\ref{truncated-delta}) permits an algebraic diagonalization of the
Hamiltonian. Each eigenvalue
$\varepsilon$ of (\ref{ho_basis}) satisfies the dispersion formula

\be\label{spectrum}
\chi_q^{-1} = -\sum^q_{n=0} {\psi^2_n(0) \over (2n+3/2) - \varepsilon/\hbar\omega} \;.
\ee
Requiring $\varepsilon=\varepsilon_{00}=\hbar \omega/2$ in (\ref{spectrum}),
we obtain a closed expression for the normalization constant
\be\label{chi}
\chi_q = - \pi^{3/2} \left(\sum^q_{n=0} {(2n-1)!! \over 2^{n} n!}\right)^{-1}\;.
\ee
We note that the sum  in (\ref{chi}) diverges as $q^{1/2}$ for large
$q$~\cite{chi}. Thus, the
strength of the $\delta$-function goes to zero as $q \rightarrow \infty$,
showing the need for a renormalization procedure.  A similar relation between the strength
of the interaction and the cutoff can be derived for
a plane-wave basis.  In that case the relation is
$V_0 = -\pi^2 \hbar^2 /\mu
\Lambda$ where $\Lambda$ is a momentum cutoff~\cite{bu06}.  This value of $V_0$
agrees with the asymptotic expression of Eq.~(\ref{chi})~\cite{chi} once we equate the corresponding cutoff energies as $\hbar^2 \Lambda^2/2 \mu = (2q+3/2) \hbar \omega$.

The excited states of the $q$-truncated Hamiltonian (\ref{ho_basis}) have energies
$\varepsilon^{(q)}_{n0}$ that differ from the exact unitary spectrum (\ref{relative-eigen}).
Using the dispersion relation (\ref{spectrum}), we find that
the error in the energy $\delta \varepsilon^{(q)}_{n0}=
\varepsilon^{(q)}_{n0} -\varepsilon_{n0}$ goes to zero at large
$q$, but only at a rather  slow rate, $\delta \varepsilon^{(q)}_{n0} \sim
q^{-1/2}$.  We present evidence below that this slow convergence
is also present in the $q$-renormalized energies for the
 $A=3$ and $A=4$ systems.  This makes it
problematic to extrapolate the $q$ series to estimate the true $q \to \infty$
energies.

\noindent{\em A new effective interaction.}
We have considerably more freedom
to construct the $q$-space interaction than we have exploited so far.
The only requirement on the $q$-space Hamiltonian is that it converge
to the unitary limit for large $q$.  For example, in
effective field theory one may introduce derivatives of
the contact interaction to fit certain properties of the two-particle
Hamiltonian.  Here we propose the following prescription to improve
the $q$-space interaction:  simply require that the relative-coordinate
Hamiltonian reproduce all $q+1$ $s$-wave eigenvalues of Eq.~(\ref{relative-eigen}).
We can do this and still keep the separable form for the interaction,
\be\label{u2}
V^{{\rm eff} (q)}_{n,n'}= -\hbar \omega f_n f_{n'}.
\ee
A motivation for preserving the separable form is given in the
discussion below. There are $q+1$ independent
variables $f_n$ in the interaction (\ref{u2}) and the same number of
eigenvalue equations having the form of Eq.~(\ref{spectrum}) with
$f_n$ replacing $\sqrt{|\chi_q|}\psi_n(0)$. Using the conditions that all
$q+1$ lowest $l=0$ unitary eigenvalues (\ref{relative-eigen}) $(n=0,\ldots,q)$ are
reproduced, we find the following $q+1$ equations for $f_n$
\be\label{f-n}
\sum^q_{n=0} {f^2_n \over 2(n-r) +1} =1 \;\;\;\;\; (r=0,\ldots,q) \;.
\ee
Eqs.~(\ref{f-n}) determine a unique solution for $f_n^2$ ($n=0,
\ldots,q$)~\cite{fail}. We choose the sign of the real numbers $f_n$ to coincide with the sign of $\psi_n(0)$. Using the convention that the harmonic oscillator wave functions be positive at the origin, the unique solution for $f_n$ is
\be\label{solution}
f_n =  \sqrt{{(2n+1)!! \over (2n)!!} {[2(q-n)-1]!! \over [2(q-n)]!!}} \;.
\ee

The interaction defined by (\ref{u2}) and (\ref{solution}) is different from the renormalized contact interaction for any $q$. However, its eigenfunction components (in the 3-D oscillator basis) converge to the corresponding unitary eigenfunction components in the limit of large $q$ with an error of $\sim q^{-1}$. In comparison, the eigenvector components of the renormalized contact interaction converge to the same unitary eigenvector components but at a slower rate of $\sim q^{-1/2}$.

\noindent{\em CI method and truncation of many-particle space.}
In the CI approach, one uses a single-particle basis in the laboratory frame
and constructs a many-particle basis of Slater determinants for $A$
fermions. In our problem, a natural choice for the single-particle basis are
the eigenstates of the three-dimensional harmonic oscillator.
These states are labeled by orbital quantum numbers $a=(n_a,l_a)$,
the orbital magnetic quantum number $m_a$, and an additional two-valued
quantum number (e.g. spin) to distinguish the two species of fermions.

A way to truncate the many-particle space must be specified,
because there is no natural
truncation associated with the interaction except
in the trivial cases $q=0$ or $A=2$.  There are a number of truncation
schemes in the literature;
here we will define a truncated single-particle orbital
basis and construct the $A$-particle wave function allowing all possible
anti-symmetrized product states.
In particular, we shall use all single-particle states in the
oscillator shells
$N=0,\ldots, N_{\rm max}$ with $N=2n_a+l_a$ to construct the
many-particle states.  There will be
two limiting processes necessary to calculate the many-particle energies.  The
first is $N_{max} \rightarrow \infty$, which we will investigate
for fixed $q$.  Then, with converged $q$-regulated energies we
estimate the $q \rightarrow \infty$ limit.

Two technical aspects of our calculations should be mentioned.
The two-particle matrix elements of the interaction in
the oscillator basis are conveniently calculated using the  Talmi-Moshinsky
brackets to transform to relative and center of mass coordinates \cite{hob}.  The many-particle Hamiltonian
is constructed and diagonalized using the nuclear shell
model code {\tt oxbash}~\cite{oxbash}.  Unlike the nuclear shell model,
our orbitals are characterized by integer angular momentum values.
The two fermion species are distinguished in the same way as
neutrons and protons are distinguished in the nuclear application.

\begin{figure}[t]
\epsfxsize= \columnwidth \centerline{\epsffile{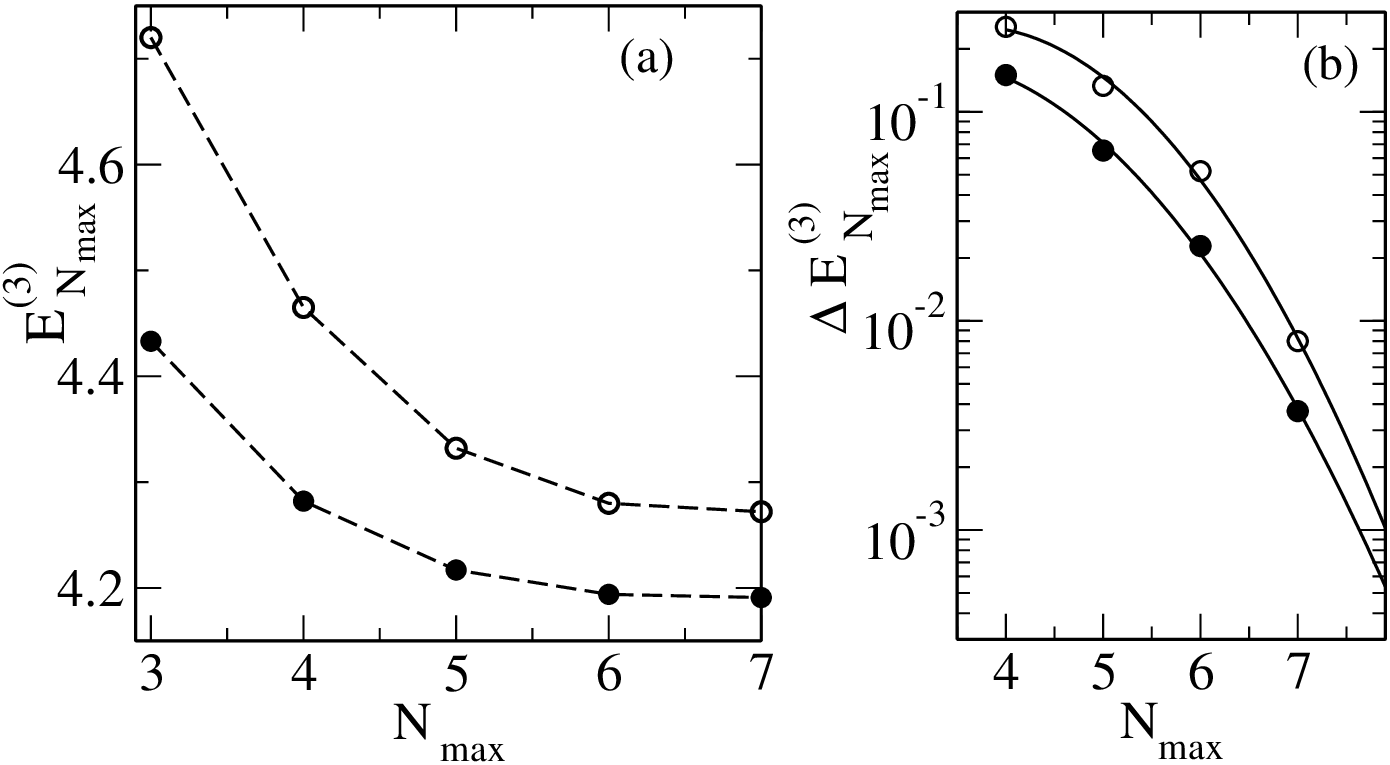}}
\caption{
Convergence in $N_{\rm max}$ for the $A=3$ ground-state energy. (a) $E^{(q)}_{N_{\rm max}}$
versus $N_{\rm max}$ for $q=3$. Open circles correspond to the renormalized
contact interaction and solid circles to the interaction defined by
(\ref{u2}) and (\ref{solution}). (b) $\Delta
E^{(3)}_{N_{\rm max}}$ versus $N_{\rm max}$ in a logarithmic scale. All energies are in units of $\hbar\omega$.}
\label{a3nmax}
\end{figure}

\noindent{\em A=3 system.}
We now show the results for $A=3$.
The ground state of the $A=3$ system is a negative-parity state with total angular momentum $L=1$
and energy $4.2727243\dots\hbar\omega$~\cite{ta04,we06}. In our CI convergence studies we
computed the ground-state energies $E^{(q)}_{N_{\rm max}}$
for $q=1,2,3,4$ and $N_{\rm max}=q,\ldots, 7$.

For a fixed $q$, we find that $E^{(q)}_{N_{\rm max}}$  converge
exponentially or better in $N_{\rm max}$ for both interactions.
This is demonstrated in Fig.~\ref{a3nmax}. Panel (a) shows
$E^{(q)}_{N_{\rm max}}$ versus $N_{\rm max}$ for $q=3$.  Both the
renormalized contact interaction (open circles) and the new interaction
(solid circles) are monotonically decreasing, as they must when the
space gets larger.  The important point, seen in Fig.~\ref{a3nmax}(b), is
that the energy differences $\Delta E^{(q)}_{N_{\rm max}} \equiv E^{(q)}_{N_{\rm max}-1} - E^{(q)}_{N_{\rm max}}$ decrease rapidly on a logarithmic scale.  In
fact, the decrease is steeper than linear on that scale, suggesting
that the convergence might be faster than exponential.
The solid lines are quadratic fits to $\log(\Delta E^{(q)}_{N_{\rm max}})$, used to extrapolate to a value of $E^{(q)}\equiv E^{(q)}_\infty$.
We observe the decrease rate of $\Delta E^{(q)}_{N_{\rm max}}$ to be
monotonically increasing with $N_{\rm max}$, so a conservative lower bound
in $E^{(q)}$ is obtained using a fixed-rate extrapolation above $N_{\rm
max}=7$ with an average rate determined by the points $N_{\rm max}=5,6,7$.
An upper bound for $E^{(q)}$ is given by $E^{(q)}_7$.

\begin{figure}[t!]
\epsfxsize= \columnwidth \centerline{\epsffile{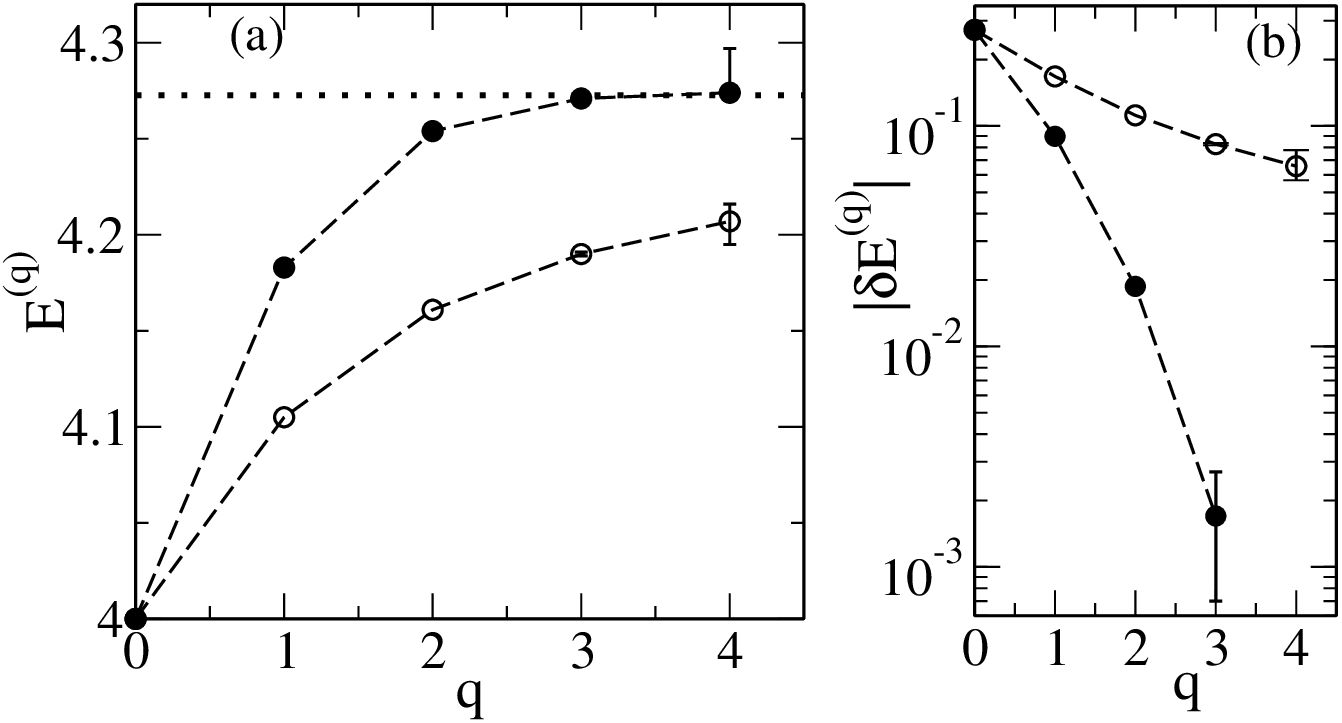}}
\caption{Convergence of the $q$-regulated energies for the $A=3$ ground state. (a) $E^{(q)}$ versus $q$ for both interactions (symbols and units as in Fig.~\ref{a3nmax}). The dotted line is the exact ground-state energy.  (b) The error $|\delta E^{(q)}|$ in a logarithmic scale.}
\label{a3nq}
\end{figure}

Fig.~\ref{a3nq}(a) shows the converged or extrapolated
energies $E^{(q)}$ versus $q$. These energies are monotonically increasing
function of $q$. For the new interaction (solid circles), we observe a fast
convergence to the known exact value (dotted line). Fig.~\ref{a3nq}(b) shows the absolute value of the deviation
$\delta E^{(q)}\equiv E^{(q)} - E^{(\infty)}$ from the exact result in a
logarithmic scale. The concavity of the curve for the renormalized contact
interaction (open circles) indicates the convergence in $q$ is slower than
exponential. We find this convergence to be consistent with a low negative
power law $\sim q^{-\alpha}$ with $\alpha$ in the range $\sim 0.5 - 1.5$ (for the excited $A=2$ system it can be shown analytically that $\alpha=1/2$). However, for the new interaction (solid circles) the convergence is at least exponential.

 This exponential convergence  allows for an accurate estimate of  $E^{(\infty)}$. We calculated successive energy differences $\Delta E^{(q)} \equiv E^{(q-1)} - E^{(q)}$ and determined an average rate of decrease $\lambda$  of $|\Delta E^{(q)}|$ for $q$ below a given $q'$.  Assuming a fixed rate $\lambda$ for $q > q'$, the extrapolated energy is $[\lambda E^{(q')} - E^{(q'-1)}]/(\lambda-1)$. We can take this value to be an upper bound for $E^{(\infty)}$, since the rate of decrease of $|\Delta E^{(q)}|$ seems to be a monotonically non-decreasing function of $q$.  Using $q'=3$ and an average decrease rate of 3.28 (determined from $\Delta E^{(q)}$ at $q=1,2,3$), we find $E^{(\infty)}= (4.274\pm 0.004)\hbar\omega$, an accuracy of $0.1\%$.

We carried out a similar study for the $L^\pi=0^+$  first excited state at
$E^{(\infty)}=4.6662\ldots\hbar\omega$~\cite{ta04,we06}. Results are shown in Fig.~3(a). As in the ground-state
case, we observe a low negative power law convergence for the renormalized
interaction and exponential convergence for our interaction. Using
$E^{(3)}$ and an average
decrease rate of 1.83 obtained from $q=1,2,3$, we estimate $E^{(\infty)}=
(4.646\pm 0.025)\hbar\omega$, an accuracy better than $0.6\%$.

\begin{figure}[t]
\epsfxsize= \columnwidth \centerline{\epsffile{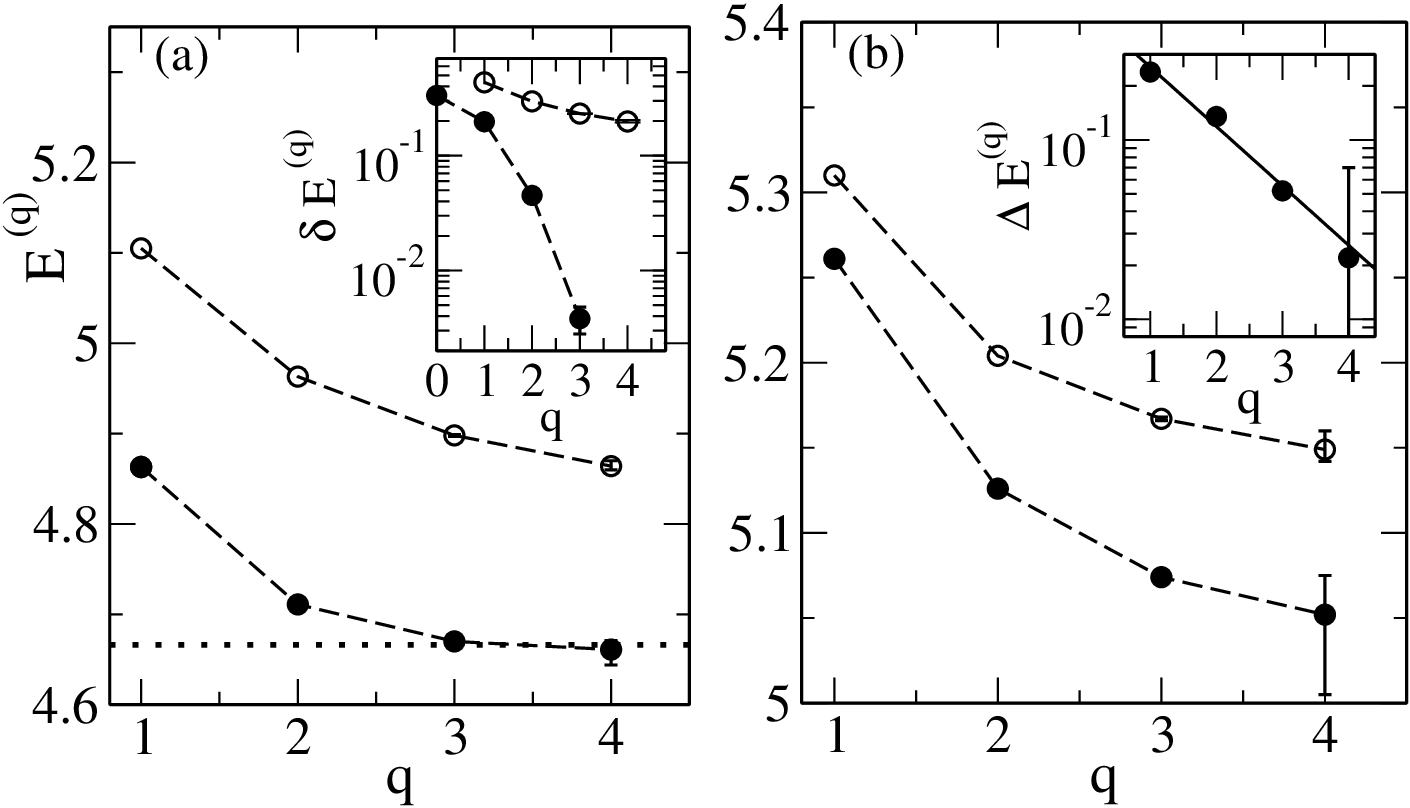}}
\caption{(a) $E^{(q)}$ versus $q$ for the lowest $L=0$ excited state of the $A=3$ system. The inset shows $\delta E^{(q)}$ versus $q$ in a logarithmic scale. Symbols and units as in Fig.~\ref{a3nq}. (b) $E^{(q)}$ versus $q$ for the $L=0$ ground state of the $A=4$ system. The inset shows $\Delta E^{(q)}$ versus $q$ for the new interaction in a logarithmic scale. The solid line is a linear fit to $q=1,2,3$.}
\label{a3pa4}
\end{figure}

\noindent{\em A=4 system.}
We also studied the $L=0$ ground state of the $A=4$ system with two particles of each species. The results for $E^{(q)}$ are shown in Fig.~\ref{a3pa4}(b). Here the exact value $E^{(\infty)}$ is unknown. An upper bound (using the new interaction) is $E^{(3)}_7= 5.075 \,\hbar\omega$. A lower bound can be obtained as for the $A=3$ system. The inset of Fig.~\ref{a3pa4}(b) shows $\Delta E^{(q)}$ in a logarithmic scale versus $q$ for the new interaction. Again, the convergence seems to be at least exponential. The straight line is a fit to $\log(\Delta E^{(q)})$ using $q=1,2,3$, and provides an average decrease rate of 2.14. Using the extrapolated $E^{(3)}=(5.074 \pm 0.001)\hbar\omega$ and this average rate, we estimate $E^{(\infty)} = (5.051 \pm 0.024)\hbar\omega$.
 Our result agrees with fixed-node Monte Carlo estimates of $(5.1+\pm 0.1)\hbar\omega$~\cite{ch07} and $(5.069\pm 0.009)\hbar\omega$~\cite{st07}.

\noindent{\em Discussion.}
There are a number of methodologies in current use to construct
effective interactions for many-particle systems; among them,
effective field theory (EFT) and the unitary-transformation method have a
connection to the interactions discussed here.  In EFT, the interaction is parameterized by contact terms (leading
order) and their derivatives.  Our procedure to construct the
$q$-renormalized contact interaction can thus be considered as
leading-order EFT.  Its poor convergence suggests
that EFT treatments will require derivative terms to
accurately model trapped fermion systems.

Our improved interaction has some connection with Suzuki's
unitary regularization~\cite{su82}, a method widely used in nuclear
physics~\cite{ba00,fu,ba06}. In Suzuki's approach, an effective interaction is
determined by a unitary transformation of the Hamiltonian that decouples a
subspace from its complementary subspace.  In practice, the
transformation is performed on the two-particle Hamiltonian, giving
a transformed Hamiltonian that is block diagonal. This block diagonal structure guarantees that the energy eigenvalues are reproduced in the truncated subspace. Our effective interaction also reproduces the exact two-particle spectrum in
a truncated subspace but has the advantage of being simple, i.e., separable.

The unitary transformation of the two-particle Hamiltonian cannot be
carried out independently for all possible pairings in the
many-body Hamiltonian. When this transformation is applied to the many-particle system, it generates higher-order many-body interactions that are
usually simply neglected.  For our Hamiltonian, additional correction terms would be required if we were to relate it to a unitary-transformed Hamiltonian. Rather than attempting to compute these correction terms, we have studied
the convergence in the large $q$-limit, where our effective interaction
coincides with the contact interaction. By studying the convergence, one can assess the usefulness of many of the specific details of the different
methodologies.  For example, there are other choices of the
many-particle space truncation that might be more efficient.  Non-unitary
transformations might give faster convergence. The
no-core-shell-model methodology~\cite{nocore} is an example where
a particular choice was made.

Our method can be applied for interaction strengths away from unitarity, at the
slight cost of inverting numerically a $(q+1)$-dimensional matrix.  It may
also be interesting to apply the method to uniform systems, using
the separability of the interaction in a plane-wave basis.
One caveat is that we have only examined three- and four-particle systems.
It will be important to confirm the exponential convergence when our interaction is used to calculate the spectra of systems with more particles.

We thank A. Bulgac, M. Forbes, M. Hjorth-Jensen and U. von Kolck for
conversations, B.A. Brown and M. Horoi for their advice on the computer
program {\tt oxbash}, and S. Fujii in particular for his guidance
on the unitary transformation models.
This work was supported in part by the U.S. DOE grants
No. FG02-00ER41132,  DE-FC02-07ER41457 and DE-FG02-91ER40608.


\begin{thebibliography}{99}
\bibitem{p2spectrum} The spectrum is most easily derived by noting that
the boundary condition $(u'/u)|_{r=0}=0$, imposed by the unitary interaction on the radial wave function $u(r)$, effectively maps the
3-D $s$-wave problem onto the even solutions of the 1-D harmonic oscillator.
\bibitem{chi} The sum in (\ref{chi}) can be performed analytically to give
$\chi_q^{-1}= -2{\pi}^{-2} \Gamma(q+\frac{3}{2})/\Gamma(q+1) \approx -2 \sqrt{\left(q + 3/4\right)}/\pi^2$.
\bibitem{fail} Fortunately the solutions $f^2_n$ are all positive; otherwise
the construction would fail.
\end{thebibliography}
\end{document}